\documentclass[citeautoscript,floatfix,aps,prb,twocolumn,superscriptaddress]{revtex4-1}
\usepackage{graphicx}
\usepackage{amsmath}
\usepackage{amssymb}
\usepackage{dcolumn}
\usepackage[version=4]{mhchem}
\usepackage{latexsym}
\usepackage{rotating}
\usepackage{epstopdf}
\usepackage[usenames,dvipsnames]{xcolor}
\usepackage{float}
\usepackage{soul}
\usepackage{epsfig}
\usepackage{psfrag}
\usepackage{natbib}
\usepackage{bm}
\usepackage{eucal}
\usepackage{mathrsfs}
\usepackage{braket}
\usepackage{enumerate}
\usepackage{longtable}
\usepackage{bm}
\usepackage{hyperref}
\usepackage{amsfonts}
\usepackage{dcolumn}
\usepackage{bm}
\usepackage{changes}

\newcommand{\be}{\begin{equation}}
\newcommand{\ee}{\end{equation}}
\newcommand{\bn}{\begin{eqnarray}}
\newcommand{\en}{\end{eqnarray}}

\def\x2y2{{x^2-y^2}}

\usepackage{color} 

\usepackage{hyperref}
\hypersetup{
colorlinks=true,final=true,
        linkcolor=red,
        citecolor=blue,
        filecolor=blue,
        urlcolor=blue,
}

\begin{document}

\title{Role of nematicity in controlling spin fluctuations and superconducting T$_{c}$ in bulk FeSe}
\author{Swagata Acharya}
 \affiliation{Institute for Molecules and Materials, Radboud University, NL-6525 AJ Nijmegen, The Netherlands}           
 \affiliation{ King's College London, Theory and Simulation of Condensed Matter,
              The Strand, WC2R 2LS London, UK}
\email{swagata.acharya@ru.nl}
\author{Dimitar Pashov}
\affiliation{ King's College London, Theory and Simulation of Condensed Matter,
              The Strand, WC2R 2LS London, UK}       
\author{Mark van Schilfgaarde}
\affiliation{ King's College London, Theory and Simulation of Condensed Matter,
              The Strand, WC2R 2LS London, UK}
\affiliation{National Renewable Energy Laboratories, Golden, CO 80401, USA}         



\begin{abstract}

FeSe undergoes a transition from a tetragonal to a slightly orthorhombic phase at 90\,K, and becomes a superconductor
below 8\,K.  The orthorhombic phase is sometimes called a nematic phase because quantum oscillation, neutron, and other
measurements detect a significant asymmetry in $x$ and $y$.  How nematicity affects superconductivity has recently
become a matter of intense speculation. Here we employ an advanced \emph{ab-initio} Green's function description of
superconductivity and show that bulk tetragonal FeSe would, in principle, superconduct with almost the same T$_{c}$ as
the nematic phase.  The mechanism driving the observed nematicity is not yet understood.  Since the present theory
underestimates it, we simulate the full nematic asymmetry by artificially enhancing the
orthorhombic distortion.  For benchmarking, we compare theoretical spin susceptibilities against experimentally observed
data over all energies and relevant momenta. When the orthorhombic distortion is adjusted to correlate with observed
nematicity in spin susceptibility, the enhanced nematicity causes spectral weight redistribution in the Fe-3d$_{xz}$ and
d$_{yz}$ orbitals, but it leads to at most 10-15$\%$ increment in T$_{c}$.  This is because the d$_{xy}$ orbital always
remains the most strongly correlated and provides most of the source of the superconducting glue.  Nematicity suppresses
the density of states at Fermi level; nevertheless T$_{c}$ increases, in contradiction to both BCS and BEC theories.  We
show how the increase is connected to the structure of the particle-particle vertex.  Our results suggest while
nematicity may be intrinsic property of bulk FeSe, is not the primary force driving the superconducting pairing.
\end{abstract}

\maketitle

\section*{Introduction}
Bulk FeSe superconducts up to 8 K, deep inside an orthorhombic phase that sets in at a much higher temperature,
90\,K~\cite{mcqueen}. The normal bulk tetragonal phase does not superconduct unless it is
doped~\cite{mizuguchi,shipra,galluzzi,sun2017,craco2014} or forms a monolayer~\cite{qing,ge2014}. A non-alloy bulk
tetragonal superconducting FeSe does not yet exist. This raises the question whether
nematicity facilitates superconductivity in the bulk or not~\cite{bohmer2017,chubukov,wang2016}. Further recent studies
have suggested that spin fluctuations are strongly anisotropic, possibly originating from electronic nematicity inside
the detwinned orthorhombic phase~\cite{toby}. This enhanced electronic anisotropy shows up in the inelastic neutron
scattering (INS), in resistivity measurements~\cite{tanatar}, in angle resolved photo-emission spectroscopic
studies~\cite{watson2016,watson2017,coldea2018} and in several other measurements~\cite{bohmer}. How such spin
fluctuations affect the superconducting instability is a subject of intensive study.

\begin{figure}[ht!]
\includegraphics[width=0.38\textwidth]{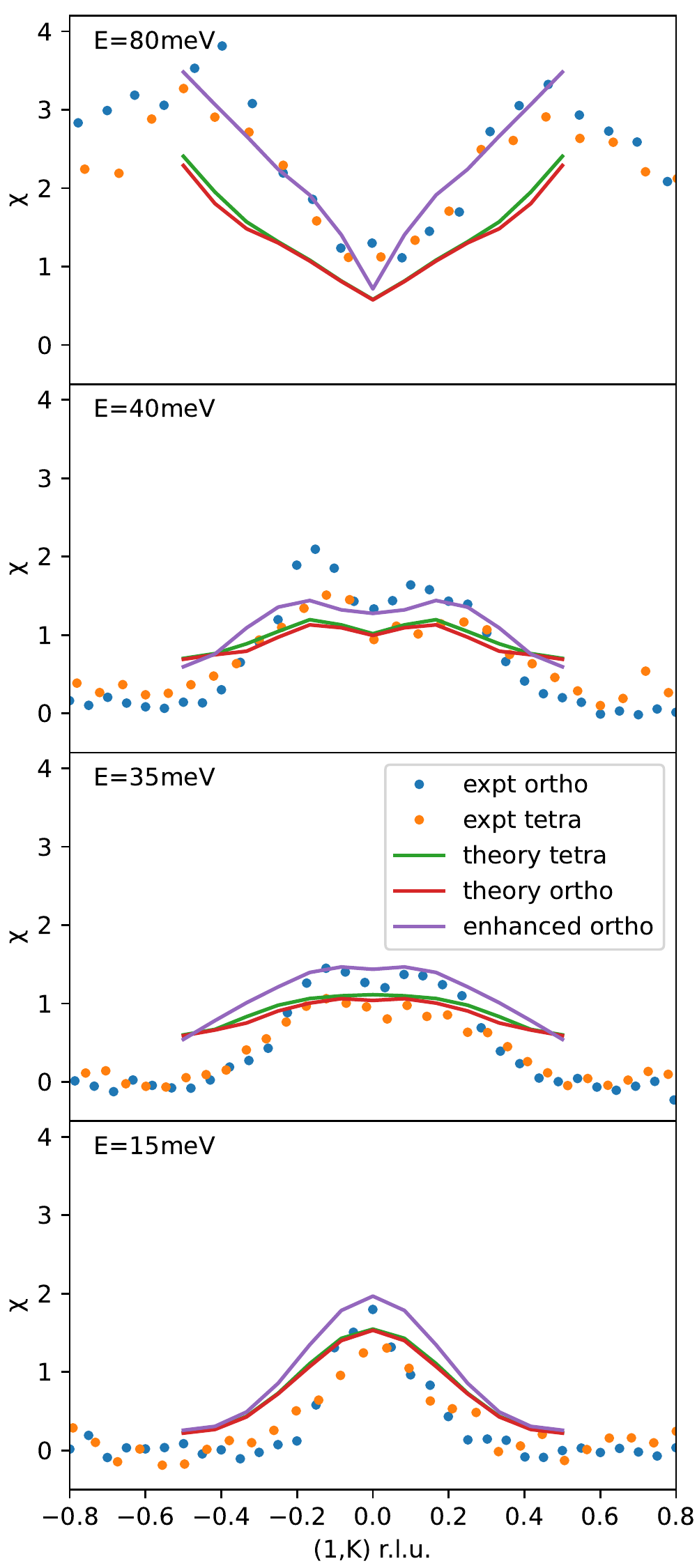}
\caption{Imaginary part of the dynamic spin susceptibility Im$\chi(q,\omega)$ plotted along the line $\bf q$=(H=1,K,L=0) in reciprocal lattice (r.l.u) units of one Fe-atom unit cell. (H,K,L) is the standard notation used to identify reciprocal lattice directions. Experimentally it is well established  that the low energy peak in Im$\chi(q,\omega)$ is at $\bf q$=(1,0,0) in the same notation. Different energy cuts are taken and plotted against the
observed Im$\chi(q,\omega)$ in inelastic Neutron scattering (INS) experiment~\cite{wang}. The experimental data is
reproduced with the raw data received from Wang et al. We plot our computed Im$\chi(q,\omega)$ at $\omega$=15, 35, 40
and 80 meV against INS data. Theoretically, the data for the tetragonal (green continuous line) and orthorhombic (red
continuous line) phases are nearly indistinguishable and both of them deviate from the experimental data (the latter is plotted as
orange (tetra) and blue (ortho) circular points). The theoretical data for the enhanced orthorhombic distortion (purple continuous line)
remarkably agrees with the experimental INS data for all energies and momentum.
}
\label{imchi}
\end{figure}

Here we take an \emph{ab initio} Green's function approach to compare spin fluctuations and superconductivity in the orthorhombic phase with the tetragonal
  phase of bulk FeSe.  This enables us to directly  probe what role nematicity
plays in governing these observed properties. 
To validate the theory we perform rigorous benchmarking against existing susceptibility data from INS
measurements~\cite{wang}.  Using the as-given orthorhombic structure, we show that spin
susceptibilities are nearly indistinguishable for the two phases. To mimic the true nematicity, we
  enhance the small observed orthorhombic distortion by a factor of five, to simulate the best
possible agreement against the existing INS data inside the orthorhombic phase.  While this enhanced structural
nematicity has a different physical origin (most likely originating from a \emph{k}
  dependent vertex in the spin channel, or in the electron-phonon interaction) it provides a similar
  contribution to nematicity by adding an effective potential from a different source.
As we show below, it enhances nematicity and generates a spin
susceptibility that fairly well replicates nematicity observed in neutron
measurements.  We can then assess its impact on superconductivity, since spin fluctuations are the primary driving force
for superconductivity, and we find that enhanced nematicity has only very moderate effect.

\section*{Methods}
We believe findings are conclusive because they are obtained from a high fidelity, \emph{ab initio} description of spin
fluctuations and superconductivity that depends only minimally on model assumptions.  Our theory couples
(quasi-particle) self consistent \emph{GW} (QS\emph{GW}) with dynamical mean field theory
(DMFT)~\cite{nickel,swag18,swag19,BaldiniPNAS}.  Merging these two state-of-the-art methods captures the effect of both
strong local dynamic spin fluctuations (captured well in DMFT), and non-local dynamic correlation~\cite{tomc,
  questaal-paper} effects captured by QS\emph{GW}~\cite{kotani}.  On top of the DMFT self-energy, charge and spin
susceptibilities are obtained from vertex functions computed from the two-particle Green's function generated in DMFT,
via the solutions of non-local Bethe Salpeter equation (BSE).  Additionally, we compute the particle-particle vertex
functions and solve the ladder particle-particle BSE~\cite{hyowon_thesis,yin,swag19} to compute the superconducting
susceptibilities and eigenvalues of superconducting gap instabilities.

Single particle calculations (LDA, and energy band calculations with the static quasiparticlized QS\emph{GW} self-energy
$\Sigma^{0}(k)$) were performed on a 16$\times$16$\times$16 \emph{k}-mesh while the (relatively smooth) dynamical
self-energy $\Sigma(k)$ was constructed using a 8$\times$8$\times$8 \emph{k}-mesh and $\Sigma^{0}$(k)
is extracted from it.  The same mesh is used for DFT.  We use Questaal's all-electron implementation
for all the calculations here; it is explained in detail in a recent methods paper \cite{questaal-paper}. For LDA we use
a Barth-Hedin exchange-correlation functional (QS\emph{GW} does not depend on the LDA, but we nevertheless
  present LDA results to show how QS\emph{GW} incorporates missing correlations from LDA and renormalizes the electronic bands further.)  The QS\emph{GW} approximation, including the all-electron product basis used to make the polarizability and self-energy is described in
detail in Ref.~\onlinecite{kotani}. Our one-particle basis set was constructed of 110 orbitals, including \emph{spdfspd}
orbitals centered on Fe augmented with local orbitals on the 4d, and \emph{spdfspd} on
Se.  The product basis in the augmentation spheres was expanded to $l=8$, and included 520 orbitals; for the Coulomb
interaction in the interstitial a plane wave cutoff of 2.3\,Ry was used.  The polarizability is computed with the
tetrahedron method; for the self-energy a smearing of 0.003 Ry was used.  For frequency integration, we used a mesh with
0.01 Ry spacing at low energy, gradually increasing at higher energy.  Six points were used on the imaginary axis
contribution to the self-energy.  The charge density was made self-consistent though iteration in the QS\emph{GW}
self-consistency cycle: it was iterated until the root mean square change in $\Sigma^{0}$ reached
10$^{-5}$\,Ry.  Thus the calculation was self-consistent in both $\Sigma^{0}(k)$ and the density.  At the end of
QS\emph{GW} cycles, we use the quasi-particlized electronic band structures as the starting point of our DMFT
calculations. The impurity Hamiltonian is solved with continuous time Quantum Monte Carlo
solver~\cite{haule,werner}. For projectors onto the Fe d subspace, we used projectors onto augmentation spheres,
following the method described in this reference~\cite{haule1}. DMFT is solved for all five Fe-3d orbitals using 
Continuous Time Quantum Monte Carlo (CTQMC)
on a rotationally invariant Coulomb interaction. The double counting correlations are implemented using fully localized
limit approximation.  The DMFT for the dynamical self energy is iterated, and converges in 30 iterations. Calculations
for the single particle response functions are performed with 10$^{9}$ QMC steps per core and the statistics is averaged
over 128 cores. The two particle Green’s functions are sampled over a larger number of cores (40000-50000) to improve
the statistical error bars.  The local effective interactions for the correlated impurity Hamiltonian are given by
\emph{U} and \emph{J}. These are calculated within the constrained RPA~\cite{ferdi} from the QS\emph{GW} Hamiltonian
using an approach~\cite{swag19} similar to that of Ref~\onlinecite{christoph}, but using projectors from
  Ref.~\onlinecite{haule1}. For bulk FeSe we find \emph{U}=3.5 eV and \emph{J}=0.6 eV from our constrained RPA
calculations. $\frac{\emph{J}}{U}$=0.17 suggests that the system is in strong Hund's metallic limit as we have discussed
in a prior work~\cite{symmetry2021}.

\begin{figure*}[ht!]
\includegraphics[width=1.0\textwidth]{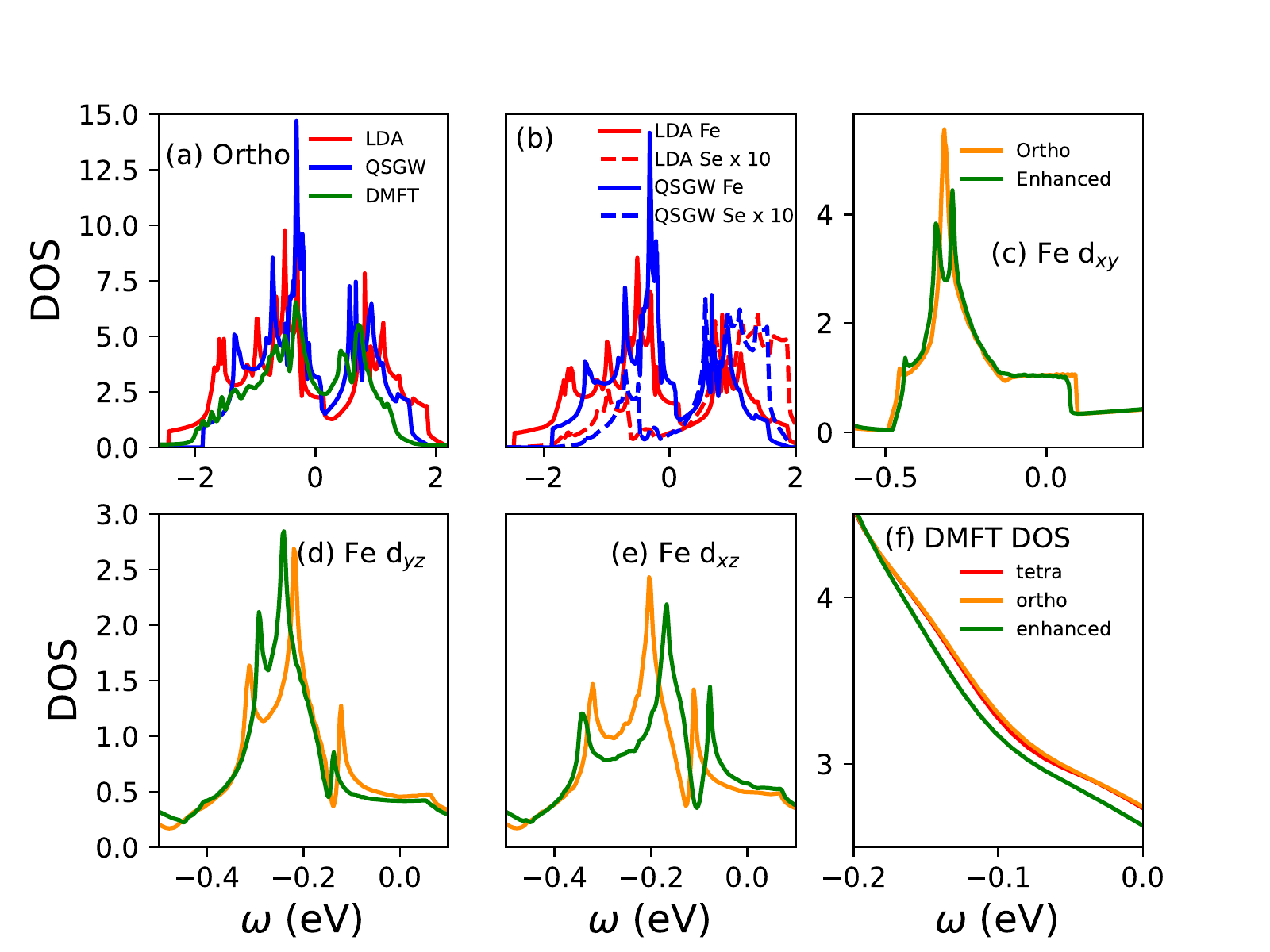}
	\caption{(a) QS\emph{GW} total local density of states (DOS) for FeSe is narrower compared to LDA. DMFT narrows the bands compared to QS\emph{GW}.(b) Both Fe-3d states and Se-p states get narrowed within QS\emph{GW}, although the Se-p states are negligibly small around $E_{F}$. (c) The artificially enhanced nematic distortion reduces the Fe-d$_{xy}$ band width weakly compared to the original nematic phase, however, its effect on (d) d$_{yz}$ and (e) d$_{xz}$ orbitals are opposite, in one case DOS at $E_{F}$ drops and in the other case it enhances. (f)  The total QS\emph{GW}+DMFT local density of states (DOS) for FeSe in tetragonal, orthorhombic  and orthorhombic phase with enhanced distortion are shown.}
	\label{dos}
\end{figure*}

\section*{Results}
\subsection*{Computational details for magnetic susceptibility} 

 We compute the local polarization bubble from the local single-particle
	Green's function computed from DMFT. We extract $\Gamma_{loc}^{irr}$, by solving the local
	Bethe-Salpeter equation which connects
	the local two-particle Green's function ($\chi_{loc}$) sampled by
	CTQMC, with both the local polarization function ($\chi_{loc}^{0}$)
	and $\Gamma_{loc}^{irr}$.
	\begin{equation}
		\Gamma_{loc{\alpha_{1},\alpha_{2}\atop \alpha_{3},\alpha_{4}}}^{irr,m(d)}(i\nu,i\nu^{\prime})_{i\omega}=[(\chi_{loc}^{0})_{i\omega}^{-1}-\chi_{loc}^{m(d)-1}]_{{\alpha_{1},\alpha_{2}\atop \alpha_{3},\alpha_{4}}}(i\nu,i\nu^{\prime})_{i\omega}.
	\end{equation}
$\Gamma$ is the local irreducible two-particle vertex functions computed in
	magnetic (m) and density (d) channels.  $\Gamma$ is a function of two fermionic frequencies $\nu$ and $\nu'$ and the bosonic
	frequency $\omega$.
	
	
Spin ($\chi^{m}$) and charge ($\chi^{d}$) susceptibilities are computed by solving the momentum
          dependent Bethe-Salpeter equations in magnetic (spin) and density (charge) channels by dressing the non-local
          polarization bubble $\chi^0$ with local irreducible vertex functions $\Gamma$ in their respective channels:
	
	\begin{equation}
		\chi_{{\alpha_{1},\alpha_{2}\atop \alpha_{3},\alpha_{4}}}^{m(d)}(i\nu,i\nu^{\prime})_{\textbf{q},i\omega}=
		[(\chi^{0})_{\textbf{q},i\omega}^{-1}-\Gamma_{loc}^{irr,m(d)}]_{{\alpha_{1},\alpha_{2}\atop \alpha_{3},\alpha_{4}}}^{-1}(i\nu,i\nu^{\prime})_{\textbf{q},i\omega}.
		\label{eq:BSE_imag}
	\end{equation}

$\chi^0$ is computed from single-particle DMFT Green's functions embedded
          in a QS\emph{GW} bath.  Susceptibilities $\chi^{m(d)}(\textbf{q},i\omega)$ are computed by summing over
          frequencies ($i\nu$,$i\nu^{\prime}$) and orbitals ($\alpha_{1,2}$).


\subsection*{Benchmarking of magnetic susceptibilities against experimental observations}

In Fig.~\ref{imchi} we plot the imaginary part of Im$\,\chi(q,\omega)$ along the $q$=(1,$K$) line of the one-atom
Brillouin zone Im$\,\chi(q,\omega)$ is computed in tetragonal and nematic phases and compared against the inelastic
neutron scattering (INS) data received from Wang et al~\cite{wang}.  Im$\chi(q,\omega)$ is plotted for
slices $\omega$ = 15, 35, 40 and 80 meV. The ratio $a/b$ in the orthorhombic phase (\emph{a}=5.3100\,\AA,
\emph{b}=5.3344\,\AA) differs by only 0.4\% from unity and the area is only slightly smaller than the tetragonal phase
\emph{a}=3.779\,\AA, $\sqrt{2}a$=5.3443\,\AA). Consequently, Im$\chi(q,\omega)$ changes little between tetragonal and
orthorhombic phases.  Most remarkably, the entire momentum dependence of Im$\chi(q,\omega)$ is rather
well reproduced for the tetragonal phase at all energies.  This is a testimony to the fact that the essential elements
that are required to produce the momentum and energy structures for Im$\chi(q,\omega)$ are already present in
the three-tier QS\emph{GW}+DMFT+BSE approximation. Nevertheless, our computed
Im$\chi(q,\omega)$ does not adequately reflect the effect of nematicity as is apparent in deviations of our theoretical
data (red solid lines) from the experimental curves.

These calculations show that nematicity probably originates from a momentum dependent self energy which contains a
longer range vertex beyond the single site approximation in DMFT.  Our DMFT spin vertex is local and also we do not have
the ability to include the electron-phonon interaction \emph{ab initio}; one of which is likely to be responsible for
the enhancement of the nematicity.  We, instead, mimic the effective potential originating from either
  of these sources by modifying the crystal-field splitting, enlarging
the orthorhombic splitting by a factor of five to (a=5.3100\,\AA, b=5.4344\,\AA).  In the rest of the paper
  we use `ortho-enhanced' to identify this particular structure.  We find that the resulting Im$\chi(q,\omega)$ is
  enhanced in intensity and produces very good agreement with the INS data over all energies and
  momentum.

The precise nature of the boson that Cooper pairs in FeSe is debated,
though generally the primary mechanism is believed to be magnetic
  fluctuations. This is supported by the observation of negligible isotope effect in bulk FeSe~\cite{isotope}.  Since
  the normal phase just above {T}$_{c}$=8\,K is orthorhombic, there has been much speculation
    that nematicity can be the mechanism driving superconductivity, though heavily debated.
This is the question we consider within the QS\emph{GW}+DMFT theory.  While it is conceivable that
    our fictitious inducement of nematicity does not yield the proper modification of superconductivity, the true
    mechanism would have to occur via some unknown process that does not involve the spin susceptibility (which we
    adequately reproduce, as we have shown).  Thus we believe modifying the effective potential via enhancing the
    orthorhombic distortion is a sufficient proxy to reliably pursue this question.

\subsection*{Effect of the nematic distortion of density of states}

First, we study the effects of nematicity on the local
density of states (DOS). The primary effect of non-local charge correlations
  included within QS\emph{GW} is to reduce the band width (see Fig.~\ref{dos}(a,b)) of FeSe compared to LDA.  We show results only for the orthorhombic phase of FeSe, but it
  is true irrespective of the structural phase considered.  The effect nematicity on d$_{yz}$ and d$_{xz}$ are mirror
  images of each other (Fig.~\ref{dos}(d,e)); this includes the change in DOS at $E_{F}$, $\rho(E_{F})$. However, the enhanced orthorhombic distortion only weakly reduces
  the d$_{xy}$ kinetic energy (Fig.~\ref{dos}(c)) and does not alter the $xy$ contribution to
    $\rho(E_{F})$.  The Se-p states remain negligibly small within an energy range of $\pm$1 eV around $E_{F}$ in all
  cases.  In Fig.~\ref{dos}(b) the Se-p DOS is shown scaled by a factor of 10 to make it visible. The effect of such distortions are negligible on the Se-p states as well and 
these states remain irrelevant for the low energy physics of FeSe.

Fig.~\ref{dos}(f) plots the QS\emph{GW}+DMFT DOS for the three systems; tetragonal, orthorhombic and
ortho-enhanced. Note that the QS\emph{GW} Green's functions are renormalized in the presence of the self-consistent single-site
DMFT self energy. For all phases there is a significant drop in
$\rho(E)$ at $\omega=0$, and it is also seen that enhanced nematicity modestly suppresses $\rho(E_{F})$ (i.e. $\omega{=}0$). 
This observation suggests that photo-emission spectroscopy should be able to see this
drop in the local DOS at low energies in systems where nematicity plays a major role, for example, in the de-twinned
sample of FeSe~\cite{toby, watson2017,rhodes2020}.  It suggests that if the superconductivity is modified by enhanced
nematicity, it does not result from a purely Fermi surface nesting driven mechanism.  Within a purely BCS picture such a
dip in $\rho(E_{F})$ would produce exponentially weak suppression of superconducting T$_{c}$.  Even while nematicity
reduces $\rho(E_{F})$, it has the opposite effect on T$_{c}$. As we will show in the following discussion, T$_{c}$
increases as a consequence of the change in two-particle vertex which promotes the superconducting pairing glue.

\begin{figure*}[ht!] 
	\includegraphics[width=0.96\textwidth]{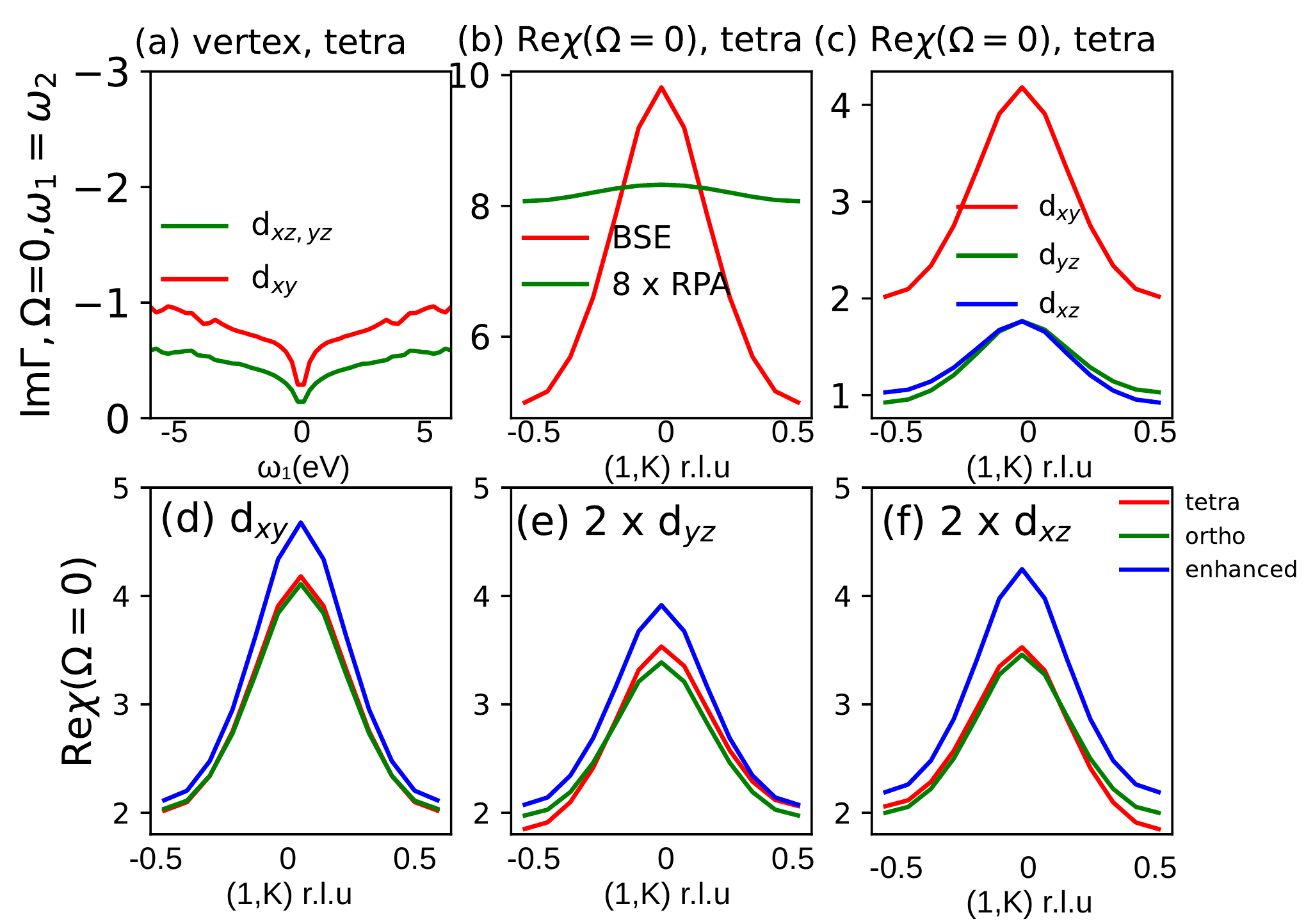} \caption{(a) Magnetic vertex and its orbital structure. For all energies, $\Gamma_{xy} {>} \Gamma_{xz,yz}$. (b) The BSE approximation is shown to strongly modify the random-phase approximation (RPA) in the magnetic channel, in a momentum dependent
          way (note that $\chi^\mathrm{RPA}$ is scaled by a factor of 8). (c,d,e,f) Orbitally resolved static spin susceptibility, Re$\chi(q)$ is plotted along \textbf{Q}=(H,$K$,L=0). The enhanced orthorhombic distortion
          increases the intensity of Re$\chi(q)$ in all intra-orbital channels, however, the d$_{xy}$ states remain
          dominant in all cases. $\chi_{xz}$ and $\chi_{yz}$ are scaled by a factor of two to improve
          visibility.}  \label{rechi}
\end{figure*}

\subsection*{Orbitally resolved Magnetic susceptibilities}

Before moving on to the discussion of superconductivity, we analyze our computed spin susceptibilities
  $\chi(q,\omega)$, particularly in an attempt to understand its orbital-structure. We show in Fig.~\ref{rechi} (a) that the irreducible vertex computed using CTQMC+DMFT is strongly orbital-dependent. It is crucially
    important that CTQMC+DMFT is site-local but not point-local. One of the
  primary successes of DMFT is its ability to pick up orbital dependent structures in self-energy and higher-order
  vertex functions. This becomes even more crucial in Hund's metals since Hund's \emph{J} generates strong
  orbital-differentiation. The local irreducible vertex computed in the magnetic channel depends on three frequencies
  (two Matsubara Fermionic frequency indices $\omega_{1,2}$ and one Matsubara bosonic frequency index $\Omega$). We show
  that for all energies (Matsubara Fermionic frequencies) the magnetic vertex $\Gamma$ remains larger in the d$_{xy}$
  channel compared to the d$_{xz,yz}$ channels, suggesting that magnetic scatterings are largest there.

The site-local vertex has a pronounced effect on the both the magnitude and momentum dependence of the
  nonlocal $\chi$ (compare RPA to BSE in Fig.~\ref{rechi}(b)), and it is the vertex that ensures that the
  antiferromagnetic instability suppresses magnetic instabilities at other $q$.
  $\chi^\mathrm{RPA}$ must be scaled by 8 or so to put on the same scale as $\chi^\mathrm{BSE}$, which
    gives a rough measure of the Stoner enhancement.  But it is important to stress once again that the orbital- and
    frequency- dependence of the vertex is crucial, and it cannot be adequately modeled by a constant.


We resolve the computed static spin susceptibilities $\chi(q)$ in different intra-orbital channels. Enhanced nematicity
causes $\chi(q)$ to increasein all intra-orbital channels; however, it remains smaller in the d$_{xz}$ and
d$_{yz}$ channels than the d$_{xy}$ channel (see Fig.~\ref{rechi} (c,d,e,f)). Orthorhombic distortion lifts
the degeneracy between the d$_{xz}$ and d$_{yz}$ orbitals, that is observed in neutron measurements. In any case, the
d$_{xy}$ channel remains the dominant spin fluctuation channel, while the d$_{xz}$ and d$_{yz}$ channels each contribute
nearly 40$\%$ of the d$_{xy}$ spin fluctuations. This is crucial. Although this desired `nematic' distortions that
produce the agreements with experimentally observed $\chi(q,\omega)$, leads to moderate spectral weight redistribution
between d$_{xz}$ and d$_{yz}$ one-particle channel, d$_{xy}$ still remains the most correlated orbital that hosts the
largest fraction of the magnetic collective fluctuations in the system.

\subsection*{Computational details of superconducting instabilities}

How particular orbitals govern spin fluctuations and thus control T$_{c}$, is key to understanding the superconducting
mechanism in such a complex multi-band manifold.  We probe the effect of such enhanced nematicity on the superconductivity. The superconducting pairing susceptibility $\chi^{p-p}$ is computed by dressing the non-local pairing
polarization bubble $\chi^{0,p-p}(\textbf{k},i\nu)$ with the pairing vertex $\Gamma^{irr,p-p}$ using the
Bethe-Salpeter equation in the particle--particle channel.

\begin{eqnarray}
	\chi^{p-p} = \chi^{0,p-p}\cdot[\mathbf{1}+\Gamma^{irr,p-p}\cdot\chi^{0,p-p}]^{-1}
\end{eqnarray}

The particle--particle vertex in the singlet channel has odd-symmetry under exchange of two external spins,

\begin{eqnarray}
	\Gamma^{p-p,s} = \frac{1}{2} [\Gamma_{\uparrow\downarrow\atop \downarrow\uparrow}^{p-p} - \Gamma_{\downarrow\uparrow\atop \downarrow\uparrow}^{p-p}]
\end{eqnarray}

The irreducible particle--particle vertex function channel $\Gamma^{p-p,irr}$ which provides the pairing glue to form
Cooper pairs, consists of the fully irreducible vertex function $\Gamma^{f,irr}$ and the reducible vertex functions computed in the particle-hole channels

\begin{eqnarray}
	\widetilde{\Gamma}^{p-h} = \Gamma^{full, p-h} -  \Gamma^{irr, p-h}
\end{eqnarray}

where $\Gamma^{full}$ is,

\begin{eqnarray}
	\Gamma^{full} =  \Gamma^{irr} -  \Gamma^{irr}\chi \Gamma^{irr}
\end{eqnarray}

This results in $\widetilde{\Gamma}^{p-h}$,

\begin{eqnarray}
	\widetilde{\Gamma}^{p-h} = \Gamma^{irr}\chi \Gamma^{irr}
\end{eqnarray}

This is one of the most crucial points of our implementation. Note that the $\Gamma^{f,irr}$ is local within
the single-site DMFT approximation and, hence, can not contribute to
superconductivity. Nevertheless, the reducible magnetic/charge vertex $\widetilde\Gamma^{irr, p-h}$, obtained from
dressing $\Gamma^{irr}$ with the full non-local and dynamic magnetic/charge susceptibilities, can have both
momentum dependence and dynamics desired for superconductivity. The Parquet like equations that are solved
 to achieve this are presented below:

\begin{eqnarray}
	&\widetilde{\Gamma}_{\alpha_{2},\alpha_{4}\atop \alpha_{1},\alpha_{3}}^{p-h,m/d}(i\nu',i\nu)_{{\bf q},i\omega} = \sum_{i\nu1,i\nu2}\sum_{{\alpha_{2}',\alpha_{4}'\atop \alpha_{3}',\alpha_{1}'}}\\
	&{\Gamma_{loc}^{irr,p-h,m/d}}_{{\alpha_{2},\alpha_{2}'\atop \alpha_{1},\alpha_{1}'}}(i\nu,i\nu1)_{i\omega}{\chi_{\bf q}^{p-h,m/d}}_{{\alpha_{2}',\alpha_{4}'\atop \alpha_{1}',\alpha_{3}'}}(i\nu1,i\nu2)_{i\omega}\nonumber\\
	&{\Gamma_{loc}^{irr,p-h,m/d}}_{{\alpha_{4}',\alpha_{4}\atop \alpha_{3},'\alpha_{3}}}(i\nu2,i\nu')_{i\omega}
\end{eqnarray}

The irreducible particle--particle vertex function $\Gamma^{irr, p-p}$ is finally written in terms of the reducible magnetic/charge vertex $\widetilde\Gamma^{m/d}$ functions.

\begin{eqnarray} 
	&\Gamma_{{\alpha_{2}\uparrow,\alpha_{4}\downarrow\atop \alpha_{1}\uparrow,\alpha_{3}\downarrow}}^{irr,p-p}(\textbf{k},i\nu,\textbf{k}',i\nu') = \Gamma_{{\alpha_{2}\uparrow,\alpha_{4}\downarrow\atop \alpha_{1}\uparrow,\alpha_{3}\downarrow}}^{f-irr}(i\nu,i\nu')\nonumber\\
	&-\frac{1}{2}[\widetilde{\Gamma}^{p-h,(d)}\nonumber\\
	&-\widetilde{\Gamma}^{p-h,(m)}]_{{\alpha_{2},\alpha_{3}\atop \alpha_{1},\alpha_{4}}}({\textbf{k}'-\textbf{k},i\nu'-i\nu})\nonumber\\
	&+\widetilde{\Gamma}^{p-h,(m)}_{{\alpha_{4},\alpha_{3}\atop \alpha_{1},\alpha_{2}}}({-\textbf{k}'-\textbf{k},-i\nu'-i\nu})
\end{eqnarray}
\begin{eqnarray} 
	&\Gamma_{{\alpha_{2}\downarrow,\alpha_{4}\uparrow\atop \alpha_{1}\uparrow,\alpha_{3}\downarrow}}^{irr,p-p}(\textbf{k},i\nu,\textbf{k}',i\nu') = \Gamma_{{\alpha_{2}\downarrow,\alpha_{4}\uparrow\atop \alpha_{1}\downarrow,\alpha_{3}\uparrow}}^{f-irr}(i\nu,i\nu')\nonumber\\
	&-\frac{1}{2}[\widetilde{\Gamma}^{p-h,(d)}\nonumber\\
	&-\widetilde{\Gamma}^{p-h,(m)}]_{{\alpha_{4},\alpha_{3}\atop \alpha_{1},\alpha_{2}}}({-\textbf{k}'-\textbf{k},-i\nu'-i\nu})\nonumber\\
	&-\widetilde{\Gamma}^{p-h,(m)}_{{\alpha_{2},\alpha_{3}\atop \alpha_{1},\alpha_{4}}}({\textbf{k}'-\textbf{k},i\nu'-i\nu})
\end{eqnarray}

Finally, exploiting the Eqn. (6,7) and Eqn. (10,11) we obtain the $\Gamma^{irr,p-p}$ in the singlet channel from the magnetic  and density  particle-hole reducible vertices,
\begin{eqnarray} 
	&\Gamma_{{\alpha_{2},\alpha_{4}\atop \alpha_{1},\alpha_{3}}}^{irr,p-p,s}(\textbf{k},i\nu,\textbf{k}',i\nu') = \Gamma_{{\alpha_{2},\alpha_{4}\atop \alpha_{1},\alpha_{3}}}^{f-irr}(i\nu,i\nu')\nonumber\\
	&+\frac{1}{2}[\frac{3}{2}\widetilde{\Gamma}^{p-h,(m)}\nonumber\\
	&-\frac{1}{2}\widetilde{\Gamma}^{p-h,(d)}]_{{\alpha_{2},\alpha_{3}\atop \alpha_{1},\alpha_{4}}}(i\nu,-i\nu')_{\textbf{k}'-\textbf{k},i\nu'-i\nu}\nonumber\\
	&+\frac{1}{2}[\frac{3}{2}\widetilde{\Gamma}^{p-h,(m)}\nonumber\\
	&-\frac{1}{2}\widetilde{\Gamma}^{p-h,(d)}]_{{\alpha_{4},\alpha_{3}\atop \alpha_{1},\alpha_{2}}}(i\nu,i\nu')_{-\textbf{k}'-\textbf{k},-i\nu'-i\nu}
	\label{eq:Gamma_pp_nonloc1}
\end{eqnarray}

With $\Gamma^{irr,p-p}$ in hand we can solve the p-p BSE to compute the p-p susceptibility $\chi^{p-p}$.

\begin{eqnarray}
	\chi^{p-p} = \chi^{0,p-p}\cdot[\mathbf{1}+\Gamma^{irr,p-p}\cdot\chi^{0,p-p}]^{-1}
	\label{eq:chipp}
\end{eqnarray}

The critical temperature $T_{c}$ is determined by the temperature where  $\chi^{p-p}$ diverges.
For such divergence the sufficient condition is that at least one eigenvalue of the pairing matrix  -$\Gamma^{irr,p-p} \cdot \chi^{0,p-p}$ approaches unity.
The corresponding eigenfunction represents the momentum structure of $\chi^{p-p}$. 
Hence $T_{c}$, eigenvalues $\lambda$ and eigenfunctions $\phi^{\lambda}$ associated with different superconducting gap
symmetries (in the singlet channel) can all be computed by solving the eigenvalue equation,
\begin{equation}
	\frac{T}{N_{k}}\sum_{k',i\nu^{\prime}}\sum_{\alpha_{2}\alpha_{4}\atop
          \alpha_{5},\alpha_{6}}\Gamma^{irr,p-p,s}_{{\alpha_{2},\alpha_{4}\atop
            \alpha_{1},\alpha_{3}}}(k,i\nu,k',i\nu')\cdot\chi^{0,p-p}_{{\alpha_{5},\alpha_{6}\atop
            \alpha_{2},\alpha_{4}}}(k,i\nu')\phi^{\lambda}_{\alpha_{5}\alpha_{6}} = \lambda \cdot \phi^{\lambda}_{\alpha_{5}\alpha_{6}}
	\label{eq:chi1}
\end{equation}

The gap function can be written in a symmetric and Hermitian form by
\begin{eqnarray}
	&\frac{T}{N_{k}}\sum_{k',i\nu^{\prime}}\sum_{\alpha_{2}\alpha_{4}\alpha_{5} \atop \alpha_{6}\alpha_{7}\alpha_{8}}(\chi^{0,p-p}_{{\alpha_{2},\alpha_{4}\atop \alpha_{1},\alpha_{3}}}(k,i\nu))^{1/2}\cdot\nonumber\\
	&\Gamma^{irr,p-p,s}_{{\alpha_{5},\alpha_{7}\atop \alpha_{2},\alpha_{4}}}(k,i\nu,k',i\nu')\cdot
	(\chi^{0,p-p}_{{\alpha_{6},\alpha_{8}\atop
			\alpha_{5},\alpha_{7}}}(k',i\nu'))^{1/2}\cdot\nonumber\\
	& \phi^{\lambda}_{\alpha_{6}\alpha_{8}} (k',i\nu')=\nonumber\\
	&\lambda \cdot \phi^{\lambda}_{\alpha_{1}\alpha_{3}} (k,i\nu)
\end{eqnarray}
It can be explicitly shown that the eigenvalues of the non-Hermitian gap equation are the same
as eigenvalues of the Hermitian gap equation.

Finally, $\chi^{p-p}$ can be represented in terms of eigenvalues $\lambda$ and eigenfunctions $\phi^{\lambda}$
of the Hermitian particle--particle pairing matrix.
\begin{eqnarray}
	\chi^{p-p}(k,i\nu,k',i\nu') & = & \sum_{\lambda}\frac{1}{1-\lambda}\cdot(\sqrt{\chi^{0,p-p}(k,i\nu)}\cdot\phi^{\lambda}(k,i\nu))\nonumber\\
	&\cdot(&\sqrt{\chi^{0,p-p}(k',i\nu')}\cdot\phi^{\lambda}(k',i\nu'))
	\label{eq:ppladder}
\end{eqnarray}

To solve this eigenvalue equation, the most important approximation we make is to take the static limit of
$\Gamma^{irr,p-p}$ in the bosonic frequency i$\omega$=0 (real frequency axis).  The explicit dependence on the fermion
frequencies are kept, as are all the orbital and momentum indices.

\begin{figure*}[ht!]
	\includegraphics[width=0.96\textwidth]{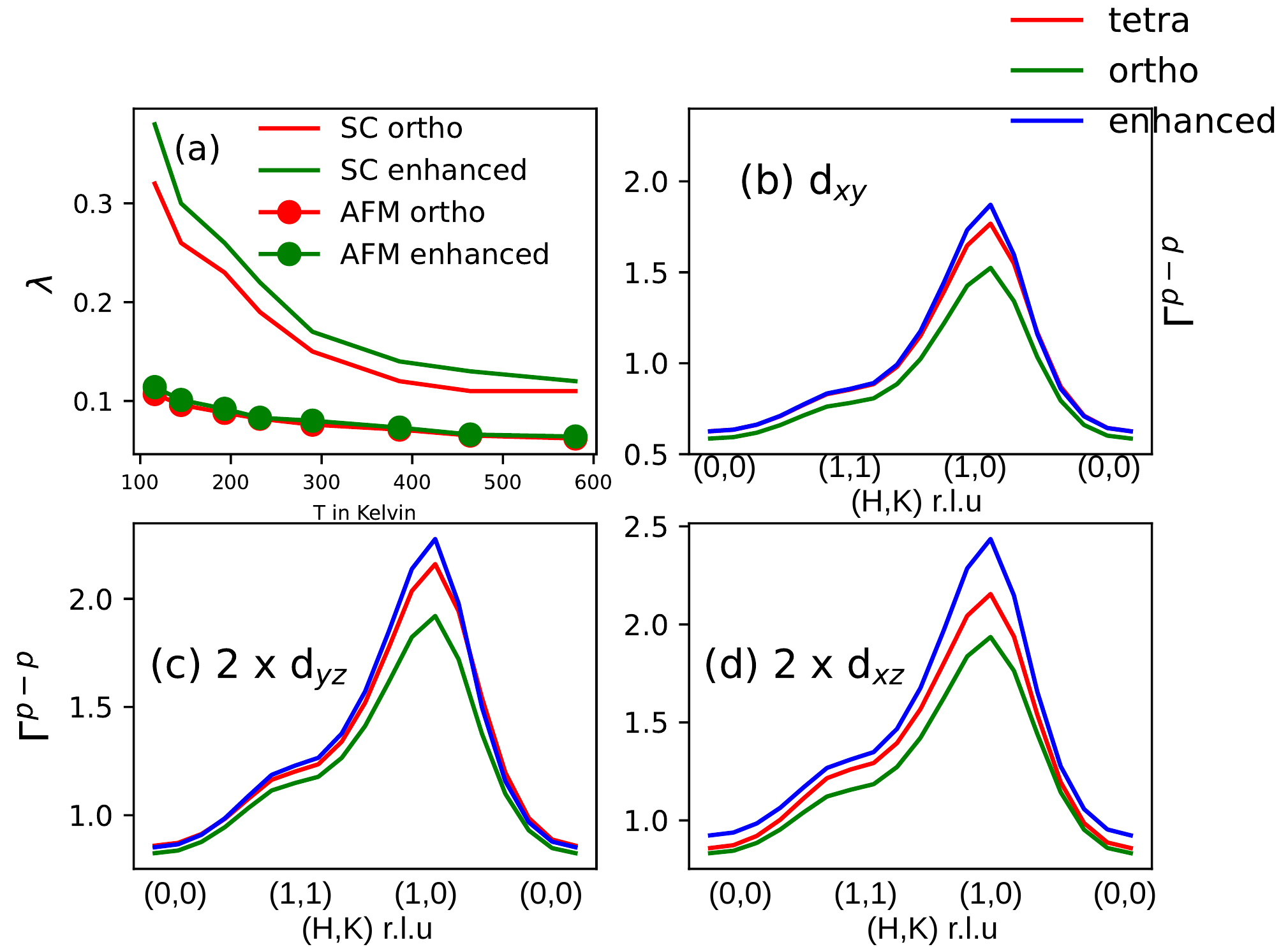}
	\caption{ (a) The leading instability in the superconducting channel is shown to dominate over the magnetic
          instability for all temperatures. In (b), (c), (d) the momentum structure of the particle-particle vertex is
          shown. The particle-particle instability remains at least twice larger in the d$_{xy}$ channel in all phases
          of FeSe compared to the d$_{yz}$ and d$_{xz}$ channels.}
	\label{vertex}
\end{figure*}

As is apparent from Equation~(\ref{eq:Gamma_pp_nonloc1}) at what wave vector spin and charge
  fluctuations are strong is of central importance to the kind of superconducting pairing symmetry they can form.  The
  entire momentum, orbital and frequency dependence of the vertex functions are computed explicitly and the BSE
  equations are solved with them. Since the vertex structure has no
  predefined form-factor, the emergent superconducting gap symmetry is calculated in an unbiased manner.  This
  provides an unbiased insight into the superconducting gap symmetries, the strength of the leading eigenvalues in
  different systems and, most importantly, allows for a fair comparison of the relative strength of the leading
  superconducting instabilities in bulk tetragonal, orthorhombic and enhanced-ortho FeSe.  Thus, our ability to predict
  these properties is limited mostly by the fidelity of the Green's functions that determine the vertices and
  $\chi$. Further, the fact that the charge component to the superconducting vertex is finite, within our
  formulation, ensures that the magnetic (in this case antiferromagnetic) instability can not drive an order and
  suppress superconductivity, as it naturally happens with density functional and other mean field approaches. We show
  in Fig.~\ref{vertex}(a) that the antiferromagnetic instability remains the sub-leading instability in FeSe making way
  for superconductivity to take place.

There is a practical limitation, however.  Since we compute the vertex functions from CTQMC, which limits the
temperatures down to which the vertex can be computed. We have observed in different materials that the leading
eigenvalue $\lambda$ does not have a simple, analytic dependence on temperature~\cite{prl2020}, and hence $\lambda$ can
not be reliably extrapolated to very low temperatures.  For that reason, we avoid estimating $T_{c}$ (the temperature at
which $\lambda$ reaches 1) for different systems from our method, rather, we compare the strength of $\lambda$ for a
given temperature in different materials (Fig.~\ref{vertex}), which is free from any ambiguities. We
have explored the fidelity of this implementation in predicting the gap symmetries and leading instabilities in several
previous works~\cite{symmetry2021,apl2021,prl2020}.

\subsection*{Leading and sub-leading superconducting instabilities}

\begin{figure}[ht!] 
\includegraphics[width=0.48\textwidth]{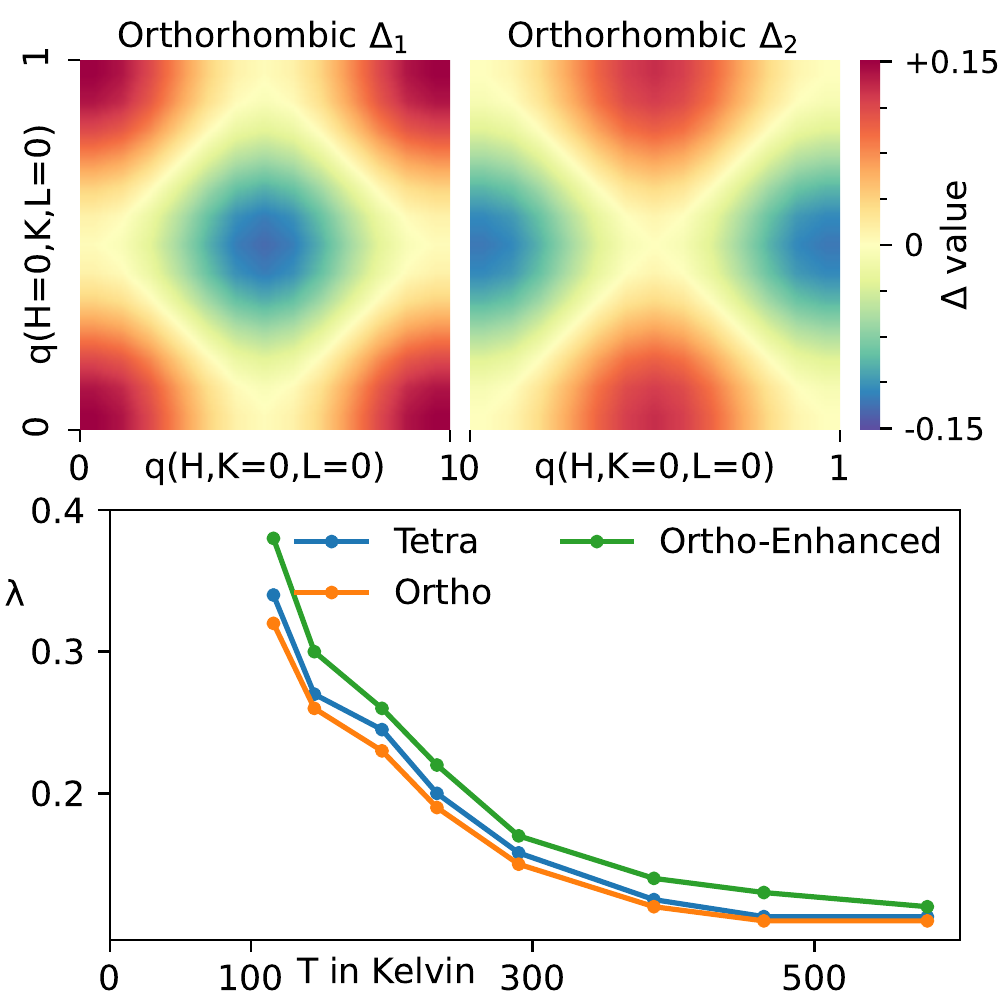}
\caption{The superconducting instabilities inside the orthorhombic phase are shown: instabilities correspond to the
  leading ($\Delta_{1}$) and and lagging ($\Delta_{2}$) eigenvalues of the solutions to the BCS gap equation. The
  evolution of the leading eigenvalue as a function of temperature is shown for tetragonal, orthorhombic and
  ortho-enhanced phases in the bottom panel.}
\label{sc}
\end{figure}

We observe that the leading eigenvalue $\lambda$ (Fig.~\ref{vertex}) of the superconducting gap
equation corresponds to an extend s-wave gap structure~\cite{mazin}, with a form factor $\Delta_{1} \sim
\cos(k_{x})+\cos(k_{y})$ (see Fig.~\ref{sc}) in all the phases.  The lagging instability $\Delta_{2}$ has a
$\cos(k_{x})-\cos(k_{y})$ structure.  Moreover, this dominant instability exists in the d$_{xy}$ orbital channel
(Fig.~\ref{vertex}) consistent with our observations of dominant spin fluctuations in the same
channel.  To see what controls the instability, it is simplest to consider what increases
  $\Gamma^{p-p}\cdot\chi^{0}$; see Eq.~\ref{eq:chipp}.  We find that the orthorhombic distortion weakly affects both
  $\chi^{0}$ (seen from the change in DOS, Fig.~\ref{dos}(f)) and $\Gamma^{p-p}$ (Fig.~\ref{vertex}) relative to
the tetragonal phase, and hence, $\Gamma^{p-p}\cdot\chi^{0}$ hardly changes. 
However, in the ortho-enhanced case, $\Gamma^{p-p}$ increases
 while $\chi^{0}$ drops. The increment in $\Gamma^{p-p}$ overcompensates the
reduction in $\chi^{0}$ so that $\Gamma^{p-p}\cdot\chi^{0}$ is
  enhanced relative to the tetragonal phase. 
We stress that the modest enhancement to T$_{c}$ occurs for reasons
completely different from approaches to
superconductivity that rely on the BCS approximation or its extension. Within BCS any changes to the
superconducting T$_{c}$ is primarily discussed with arguments based on density of states. This is natural since the
replacement of the density $\rho$ by $\rho(E_{F})$ in BCS gap integrals is strictly possible only within the BCS
approximation $E_{F} \gg \hbar\omega_{D} \gg \Delta_{0}$ (where $\omega_{D}$ is the phonon-frequency and $\Delta_{0}$ is
the superconducting gap at T=0). Based on BCS theory~\cite{bcs}, a small suppression in $\rho$(E$_{F}$) would cause
exponentially small reduction in T$_{c}$. Further, in the limit of strong electron-electron interaction (attraction) for
dilute mobile charges, T$_{c}$ is given by the Bose-Einstein condensation (BEC)
temperature~\cite{bose1924,bose19241,einstein1924}, and not by the BCS limit. In BEC T$_{c}$ has a power-law dependence
on $\rho$ in contrast to the exponential dependence in BCS. The famous Uemura plot~\cite{uemura} as was established in
the early days of cuprates, showed how unconventional superconductivity in cuprates was more akin to the BEC limit,
instead of the BCS limit. Nevertheless, in FeSe the BEC formula would again lead to weak suppression of T$_{c}$ due to
nematicity. In stark contrast, within our implementation of this finite temperature instability approach to
superconductivity, no such approximations are made, moreover, we keep full energy dynamics of the one-particle Green's
functions and full energy-dynamics (dependence on two Matsubara fermionic frequency indices and one Matsubara bosonic
frequency index) of the two-particle vertex functions $\Gamma^{p-p}$. The approximation we do make, namely put the
bosonic frequency to zero while diagonalizing the gap equation, is a sensible one that assumes that superconductivity is
a low-energy phenomena.  As noted, the full theory also predicts nematicity slightly enhances T$_{c}$,
because of changes to the vertex.

As we showed in our recent work~\cite{prl2020}, it is extremely challenging to perform the calculations at a low enough
temperature where $\lambda$ can reach unity, owing to several technicalities primarily
related to the stochastic nature of the CT-QMC solver. The lowest temperature where we could perform our calculations in
these systems is 116 K, where $\lambda$ remains less than unity. However, following the
trend, it appears that the ortho-enhanced phase can at most realize a 10-15$\%$ enhancement in T$_{c}$ in comparison to
the bulk orthorhombic phase. Our calculations suggest that the tetragonal phase could also superconduct with a T$_{c}$
not much different from the ortho-enhanced phase. This is consistent with the experimental observation that the
FeSe$_{1-x}$S$_{x}$ alloy realizes a superconducting T$_{c}$ not significantly different from the bulk orthorhombic
FeSe, even when sulfur doping suppresses the nematic phase completely~\cite{feses1,feses2,feses3}. The
  primary origin of this, as we show, is that in bulk FeSe, in all the phases, Fe-3d$_{xy}$ remains the most correlated
  orbital that sources the largest fraction of collective magnetic fluctuations and thus acts as the primary glue for the Cooper pair formation. However, as we explicitly show in
  Fig.~\ref{vertex} (b,c,d) that it is not only the magnetic susceptibilities, but the particle-particle vertex
  $\Gamma^{p-p}$ that is a complicated combination of both magnetic and charge vertex functions (see
  Eq.~\ref{eq:Gamma_pp_nonloc1}), remains most dominant in the d$_{xy}$ channel in all phases of bulk FeSe. While
  nematic distortions drive spectral weight redistribution mediated by degeneracy lifting of d$_{xz}$ and d$_{yz}$
  orbitals, they act as sub-leading channels for magnetism and superconductivity.

\section*{Conclusions}
To summarize, we perform \emph{ab initio} calculations for bulk tetragonal and orthorhombic phases of FeSe and compute
single and two-particle spectra and superconducting eigenvalues. We find that spin fluctuations are dominant in
the Fe-3d$_{xy}$ channel in all cases and can potentially drive superconductivity in the bulk tetragonal
FeSe. Nevertheless, a rigorous comparison against the observed spin susceptibilities in inelastic neutron scattering
experiments in the orthorhombic phase reveals that our computed susceptibilities have the correct momentum structure at
all energies, but not the intensity. We show that an artificially enhanced structural orthorhombic distortion simulates
the missing spin fluctuation intensity and acts as the proxy for the desired nematicity, missing from our theory but
present in the real world. This enhanced nematicity,  even while
  suppressing the one-particle density of states at the Fermi energy, nevertheless
leads to enhanced correlations from the
particle-particle superconducting correlations, leading to an increment in T$_{c}$ on the order of 10-15\%.

\section*{Acknowledgments}
We thank Qisi Wang and Jun Zhao for sharing with the raw data for spin susceptibilities. This work was supported by the Simons Many-Electron Collaboration.
SA is supported (in later stages of this work) by the ERC Synergy Grant, project 854843 FASTCORR (Ultrafast dynamics of correlated
electrons in solids).   M.v.S. and D.P. were supported by the U.S. Department of Energy, Office of Science, Basic Energy Sciences under award FWP ERW7906. We acknowledge PRACE for awarding us access to
SuperMUC at GCS@LRZ, Germany,  Irene-Rome hosted by TGCC, France, STFC Scientific Computing Department's SCARF cluster, Cambridge Tier-2 system operated by
the University of Cambridge Research Computing Service (www.hpc.cam.ac.uk) funded by EPSRC Tier-2 capital grant
EP/P020259/1.  This work was also partly carried
out on the Dutch national e-infrastructure with the support of SURF Cooperative.


%

\end{document}